\newif{\ifjournal}
  \journaltrue

\ifjournal
  \documentclass[usenatbib]{mn2e}
  \usepackage{graphicx,mathptm}
  \newcommand{\gtrsim}{\ga} 
  \newcommand{\lesssim}{\la} 
\else
  \documentclass{paper}
  \newcommand{\ga}{\gtrsim} 
  \newcommand{\la}{\lesssim} 
\fi

\def\lsim{\lesssim}

\def\ol{\Omega_{\rm de}}
\def\om{\Omega_{\rm m}}

\renewcommand{\d}{\mathrm{d}}

\begin{document}  
  
\title{Constraints on dark energy models from galaxy clusters with
  multiple arcs}
 
\ifjournal\author[Meneghetti et al.]
 {Massimo Meneghetti$^{1}$, Bhuvnesh Jain$^{2}$, Matthias
  Bartelmann$^{1}$ and Klaus Dolag$^{3}$\\
  $^1$ITA, Universit\"at Heidelberg, Tiergartenstr.~15, D--69121
  Heidelberg\\
  $^2$Department of Physics, University of Pennsylvania, Philadelphia,
  PA 19101, USA\\
  $^3$Dipartimento di Astronomia, Universit\`a di Padova, Vicolo
 dell'Osservatorio 2, I--35120 Padova}
\else
 \author{Massimo Meneghetti$^{1}$, Bhuvnesh Jain$^{2}$, Matthias
  Bartelmann$^{1}$, and Klaus Dolag$^{3}$\\
  $^1$ITA, Universit\"at Heidelberg, Tiergartenstr.~15, D--69121
  Heidelberg\\
  $^2$Department of Physics, University of Pennsylvania, Philadelphia,
  PA 19101, USA\\
  $^3$Dipartimento di Astronomia, Universit\`a di Padova, Vicolo
  dell'Osservatorio 2, I--35120 Padova}
\fi

  
\ifjournal\maketitle\fi
  
\begin{abstract}
We make an exploratory study of how well dark energy models 
can be constrained using lensed arcs at different redshifts 
behind cluster lenses. Arcs trace the critical curves of clusters, 
and the growth of critical curves with source redshift is sensitive to the
dark-energy equation of state. Using analytical models and numerically
simulated clusters, we explore the key factors involved in using
cluster arcs as a probe of dark energy. 
We quantify the sensitivity to lens mass, concentration and ellipticity
with analytical models that include the effects of dark energy on halo
structure. We show with simple examples how degeneracies between mass models
and cosmography may be broken using arcs at multiple redshifts or
additional constraints on the lens density profile.  However we
conclude that the requirements on the data are so stringent that it is
very unlikely that robust constraints can be obtained from individual 
clusters. We argue that surveys of clusters, analyzed in conjunction with 
numerical simulations, are a more promising prospect for arc-cosmography. 

We use such numerically simulated clusters to estimate how
large a sample of clusters/arcs could provide interesting constraints 
on dark energy models. We focus on the scatter produced by differences
in the mass distribution of individual clusters. We find from our
sample of simulated clusters that at least 1000 pairs of arcs are 
needed to obtain constraints if the mass distribution of individual 
clusters is taken to be undetermined. We discuss several unsolved 
problems that need study to fully develop this method for precision 
studies with future surveys. 
\end{abstract}

\ifjournal\else\maketitle\fi


\section{Introduction}

The compelling observational evidence for an accelerating expansion of
the universe has led to much recent work on models for dark energy
that would drive this expansion (see \citealt{PE02.2,PA03.1} and
\citealt{CA03.1} for reviews). Unless the dark energy is a
cosmological constant, it is likely that it evolves with time. The
best observational probes of dark energy are therefore ones that
measure the geometry of the universe at different redshifts, allowing
for a mapping of the evolution of dark energy. So far Type-Ia
supernovae, which measure the luminosity distance over a range of
redshifts, and the multipole
orders of the acoustic peaks in the power spectrum of the cosmic
microwave background, have been the only geometric probes of dark
energy (\citealt{PE99.1,RI98.1,SP03.1}; see \citealt{HU01.2} for a
discussion of different probes).

The multiple imaging of background galaxies into arcs by foreground
galaxy clusters has been used to probe the mass distribution of
clusters. The position of the arcs on the sky depends on the mass
enclosed within the lensing region of the cluster and also on ratios
of angular diameter distances, which we will refer to as the geometric
information probed by lensing. Observations of multiple arcs at
different redshifts have been suggested as probes of dark energy
\citep{LI98.1,GO02.2,SE02.1}. The idea is that the relative positions
of arcs at different redshifts depend only weakly on the mass
distribution and therefore probes the lensing geometry. Recently
\cite{SO04.1} analysed Abell 2218 using arcs at four different
redshifts to constrain dark energy models, obtaining $w\lesssim-0.85$
at 68.3\% confidence for constant $w$.

A critical issue in the reliability of geometric measurements from
cluster lensing is the sensitivity of the results to the mass
distribution of the cluster. If the cluster mass were smoothly
distributed, then it is easy to see that multiple arcs are a good
probe of geometry, but realistic clusters are likely to have
substructure and ellipticity. Ideally one would like to be able to
independently measure the mass distribution responsible for lensing,
so that no theoretical assumptions need be made in inferring geometric
information (e.g. \citealt{CH02.2}).

In this paper we use numerical simulations of lensing clusters to
explore two approaches to constraining dark energy models with cluster
arcs. The first may be called the ``golden lens'' approach: we ask how
external information about the cluster mass distribution can allow
dark energy constraints to be obtained from just one or a few lens
systems. The second is a statistical approach: we ask how large a
sample of clusters is needed to make an ``ensemble averaged''
measurement of critical curve versus redshift, which can then be
compared to simulations. The two approaches seek different kinds of
datasets, and the latter assumes that simulations (calibrated to some
extent from the data) represent lensing clusters adequately well 
at least for the giant arcs.

We describe our parameterisation of dark energy models in Sect.~2.
The lensing formalism is presented in Sect.~3, and the analytic
dependence of critical curves on cosmology is shown. In Sect.~4, we
extend the study to numerical cluster models, and we summarise in
Sect.~5.

\section{Dark-energy models\label{dem}}

We work with the metric
\begin{equation}
ds^2\, =\, a^2\left[-(1+2\phi)d\tau^2\,+\, (1-2\phi)
    (d\chi^2+r^2d\Omega^2)\right],
    \label{metric}
\end{equation}
where we have used the comoving coordinate $\chi$ and
$a(\tau)=(1+z)^{-1}$ is the scale factor as a function of conformal
time $\tau$. We adopt units such that $c=1$. The comoving angular
diameter distance $r(\chi)$ depends on the curvature: we assume a
spatially flat universe so that $r(\chi)=\chi$.  The density parameter
$\Omega$ has contributions from mass density $\om$ or dark energy
density $\ol$, so that $\Omega=\om + \ol$.

Dark energy models are often described in terms of the equation of
state $p=w\rho$, with $w=-1$ corresponding to a cosmological
constant. The time dependence of $w$ is commonly parameterised as
$w=w_0+w_a (1-a)$ \citep{LI03.1}. For comparison with other work we
will compare $w_a$ to $w'$ defined by $w=w_0+w'z$.  The Hubble
parameter $H(a)$ is given by
\begin{equation}
H(a) = H_0\left[\Omega_{m} a^{-3} + 
\Omega_{\rm de} e^{-3  \int_1^a d \ln a' (1+w(a'))}\right]^{1/2} ,
\label{hubble}
\end{equation}
where $H_0$ is the Hubble parameter today. The comoving distance
$\chi(a)$ is
\begin{equation}
\chi(a) = \int_1^a \frac{d a'}{a'^2 H(a')} ,
\label{chi} 
\end{equation}
The above equation shows how the lensing observables depend on
integrals over the dark-energy dependent expansion rate.

We consider four different cosmological models in this study (see
Table 1 and \citealt{DO03.2} for details). These can be described in
terms of an effective values of $w_0$ and $w_a$.  One is the standard
cosmological-constant model with flat geometry and
$\Omega_\Lambda=0.7$. The second has a constant ratio $w_0=-0.6$
between pressure and energy density of the dark energy (DECDM). The
remaining two use different descriptions for the self-interaction
potential of the dark-energy scalar field. One has a power-law
potential (Ratra-Peebles, RP; see \citealt{PE02.2}), the other has a
power-law potential multiplied by an exponential (SUGRA;
\citealt{BR00.2}). The RP and SUGRA models have the same values of $w$
at the present time, $w_0\simeq 0.8$, but very different evolution
histories. See \cite{DO03.2} for other details of the models and their
implementation in the simulations used here.

\begin{table}
\centering
\caption{Parameters characterising the cosmological models}
\begin{tabular}{||c|c|c|c|c|c|c||} \hline \hline
Model   &$\Omega_{\rm m}$ & $\Omega_{\rm dm}$ & $H_0/100$ & $\sigma_8$ & $w_0$ & $w_a$ \\ 
\hline \hline
$\Lambda$CDM    & 0.3 & 0.7 & 0.7 & 0.9 & -1 & 0      \\ 
DECDM   & 0.3 & 0.7 & 0.7 & 0.9 & -0.6  & 0   \\ 
RP      & 0.3 & 0.7 & 0.7 & 0.9 & -0.83 & 0.1 \\ 
SUGRA   & 0.3 & 0.7 & 0.7 & 0.9 & -0.83 & 0.5 \\ 
\hline \hline
\end{tabular}
\label{table:param}
\end{table}

\section{Analytic models}

\subsection{General properties}
\label{sect:genprop}
We adopt the density profile proposed by \cite{NA97.1} (hereafter NFW)
for modelling cluster lenses, which was found to fit numerically
simulated galaxy clusters well. We investigate here how strong lensing
by analytical NFW halos changes when the dark-energy equation of state
is changed, and postpone effects of the substructure and asymmetry of
realistic halos to Sect.~5. The profile is given by
\begin{equation}
  \rho(r)=\frac{\rho_\mathrm{s}}
  {(r/r_\mathrm{s})(1+r/r_\mathrm{s})^2}\;,
  \label{eq:nfw}
\end{equation}
where $\rho_\mathrm{s}$ and $r_\mathrm{s}$ are characteristic density
and distance scales, respectively. These two parameters are strongly
correlated \citep{NA97.1}.

The concentration is defined as $c=r_{200}/r_\mathrm{s}$, where
$r_{200}$ encloses a mean halo density of $200$ times the
\emph{critical} cosmic density. Numerical simulations show that $c$
depends on the virial mass $M$ of the halo, which can thus be used as
the only free parameter. Several algorithms have been developed for
relating $c$ to $M$. In this work, we adopt that proposed by
\citet{EK01.1} because, as recently demonstrated by \citet{DO03.2}, it
performs very well also in dark-energy cosmologies. The concentration
also depends on the cosmological model, implying that the lensing
properties of haloes with identical mass are different in different
cosmological models if they are modelled with the NFW profile. This is
an important difference of our work to earlier studies
\citep{LI98.1,GO02.2,SE02.1}.

The lensing properties of the NFW profile have been widely
investigated in the past \citep[see
e.g.][]{BA96.1,WR00.1,LI02.1,PE02.1,ME03.1}, thus we summarise them
only briefly. Lensing is fully described by the lensing potential
$\psi$. For axially symmetric models, computing the lensing potential
reduces to a one-dimensional problem. We define the optical axis as
the straight line passing through the observer and the lens centre and
introduce the physical distances perpendicular to the optical axis on
the lens and source planes, $\xi$ and $\eta$, respectively. We then
choose $r_\mathrm{s}$ as a length scale in the lens plane and define
the dimensionless distance $x\equiv\xi/r_\mathrm{s}$ from the lens
centre. By projecting $r_\mathrm{s}$ to the source distance, we define
a corresponding dimensionless distance
$y\equiv(\eta/r_\mathrm{s})(D_\mathrm{l}/D_\mathrm{s})$ from the
optical axis in the source plane.

Defining
$\kappa_\mathrm{s}\equiv\rho_\mathrm{s}r_\mathrm{s}\Sigma_\mathrm{cr}^{-1}$,
where $\Sigma_\mathrm{cr}=[c^2/(4\pi
G)]\,[D_\mathrm{s}/(D_\mathrm{l}D_\mathrm{ls})]$ is the critical
surface mass density for strong lensing and $D_\mathrm{ls}$ is the
angular diameter distance between the lens and the source planes, the
lensing potential can be written as
\begin{equation}
  \psi(x)=4\kappa_{\rm s} g(x)\:,
\label{eq:nfwPsi}
\end{equation}
where
\begin{equation}
  g(x)=\frac{1}{2}\ln^2 \frac{x}{2}+\left\{
    \begin{array}{l@{\quad \quad}l}
      2\,\mbox{arctan}^2\sqrt{\frac{x-1}{x+1}} & (x>1) \\
     -2\,\mbox{arctanh}^2\sqrt{\frac{1-x}{1+x}} & (x<1) \\
      0 & (x=1)
    \end{array}\right.\;.
\end{equation}

Axially symmetric models are generally inappropriate for describing
lensing by galaxy clusters \citep{ME03.1} because their typically high
degree of asymmetry and substructure changes their lensing properties
qualitatively and substantially. The tidal (shear) field produced by
an asymmetric mass distribution can be partially mimicked by including
ellipticity into the model. We adopt the model proposed by
\citet{ME03.1}, who obtained an elliptical generalisation of the NFW
axially symmetric lensing potential by substituting in
Eq.~(\ref{eq:nfwPsi})
\begin{equation} 
  x\rightarrow x'=\sqrt{\frac{x_1^2}{(1-e)}+x_2^2(1-e)}\;,
\end{equation}
where $x_1$ and $x_2$ are the two Cartesian components of $x$,
$x^2=x_1^2+x_2^2$. We define the ellipticity as $e\equiv1-b/a$, where
$a$ and $b$ are the major and minor axes of the ellipse, respectively.

The lensing potential implies the Jacobian matrix of the lens mapping,
\begin{equation}
  A\equiv \left(\delta_{ij}-\frac{\partial^2 \psi(\vec{x})}{\partial
  x_i \partial x_j} \right) \ , 
\label{equation:jacobian}
\end{equation}
whose determinant vanishes on the critical curves of the lens. In
particular, the tangential critical curve is located where the
tangential eigenvalue of the Jacobian matrix vanishes,
\begin{equation}
  \lambda_{\rm t}=1-\kappa-\gamma=0 \;,
  \label{eq:lambdat}
\end{equation}
where $\kappa$ and $\gamma=(\gamma_1,\gamma_2)$ are the lens
convergence and shear, respectively. They can be written in terms of
the lensing potential as
\begin{equation}
  \kappa(\vec{x})=\frac{1}{2}(\psi_{11}+\psi_{22})
  \label{eq:conv}
\end{equation}
and 
\begin{eqnarray}
  \gamma_1(\vec{x}) &=& \frac{1}{2}(\psi_{11}-\psi_{22})
	\label{equation:shear1} \\
  \gamma_2(\vec{x}) &=& \psi_{12}=\psi_{21} \ , 
	\label{equation:shear2}
\end{eqnarray}
where we abbreviate
\begin{equation}
  \frac{\partial^2 \psi(\vec{x})}{\partial x_i
    \partial x_j} \equiv \psi_{ij} \ .
\end{equation}

\subsection{Sizes of critical curves and cosmology}

\begin{figure}
  \includegraphics[width=1.0\hsize]{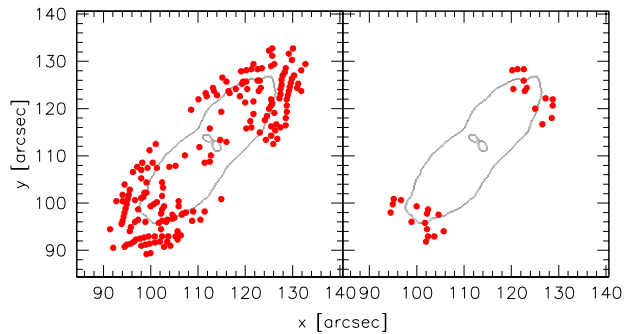}
\caption{Position of gravitational arcs with length-to-width ratio
  larger than 5 (left panel) and 10 (right panel) in a lensing
  simulation; also shown are the critical curves of the lensing
  cluster. }
\label{fig:arcpos}
\end{figure}

Tangential arcs form near the tangential critical curves of a lensing
cluster. We assume in the following that there is a correspondence
between the tangential critical curves and the observed position
of tangential arcs. In practice, tangential arcs with moderate
length-to-width ratio may form quite far from tangential critical
curves if the cluster is embedded into a strong shear field. To reduce
this uncertainty, arcs with a large tangential-to-radial magnification
ratio must be used.

The problem is illustrated in Fig.~\ref{fig:arcpos}, where we show the
position of all images with length-to-width ratio $\ge5$ (left panel)
and $\ge10$ (right panel), found in a ray-tracing simulation using a
numerical cluster model as a lens. A large number of elliptical
sources were distributed in the region containing the lens caustics in
order to obtain a large number of arcs. In this particular case, the
tangential critical curve appears very elongated, showing that the
cluster mass distribution is far from axially symmetric. From the
first panel it is clear that arcs with substantial tangential
distortion form in a wide region surrounding the critical curves. For
example, using arcs with a length-to-width ratio $\ge5$, the spread
in arc positions is $\sim10''$; selecting arcs with 
high length-to-width ratio ($\ge10$) reduces the spread to a few arc seconds.

It is important to note that arcs with large length-to-width
ratios form along those portions of the critical curves which
are at the largest distances from the cluster centre. 
In comparing theory to observations, one possibility is
to compare in detail the arcs in a simulated sample with
observed arcs. However to avoid the 
computational expense of lensing source galaxies through the numerical 
clusters discussed in the next section, we will use an average 
size of the critical curve weighted with the inverse radial
magnification as a proxy for the location of large arcs.
Hence we do the same here by assigning to the $i$-th point on the tangential
critical curve a weight:
\begin{equation}
  p_i=|2[1-\kappa(\vec x_i)]| \;.
\end{equation}
The size of the critical curve is then measured as the weighted
average distance of the critical points from the lens centre, 
\begin{equation}
  x_{\rm c}=\frac{\sum_i p_i |\vec x_i|}{\sum_i p_i} \;.
  \label{eq:critlsize}
\end{equation}
The weight factor that we have used typically underestimates the
location of giant arcs. However we use it for simplicity since we do
not expect it to significantly change the relative sizes of critical
curves at different redshifts. \cite{DA03.1} give a more detailed
discussion of the location and orientation of giant arcs in terms of
the profiles of $\kappa$ and $\gamma$.

\begin{figure}
  \includegraphics[width=\hsize]{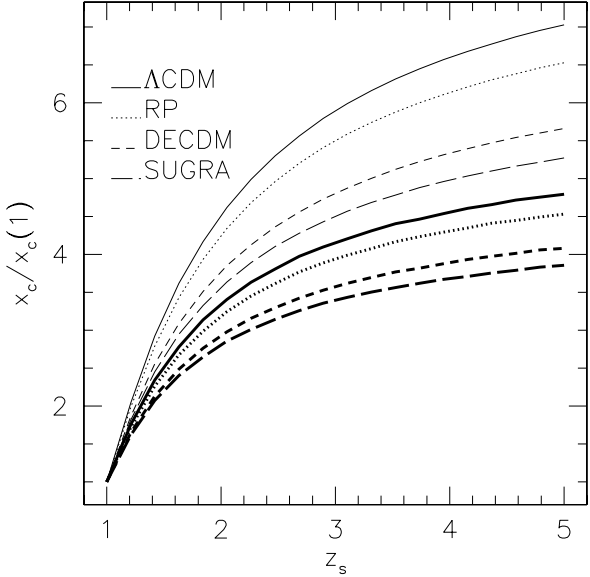}
\caption{Sizes of the critical curves for sources at
  $z_{\rm s}$ normalised to their value at $z_{\rm
  s}=1$. Results are shown for four cosmological models. The lens
  mass is $7\times 10^{14}~h^{-1}M_{\odot}$ and the lens redshift is
  $z_{\rm l}=0.6$. The lens is modelled using both an axially
  symmetric (thin curves) and a pseudo-elliptical (heavy curves) model
  with NFW density profile. In the elliptical case, the ellipticity of
  the iso-potential contours is $e=0.3$. }
\label{fig:grow}
\end{figure}

We assume here that the cluster lens is at redshift $z_{\rm l}=0.6$,
and shift the source plane from $z_{\rm s}=1$ to $z_{\rm s}=5$. The
virial mass of the lens is $7\times10^{14}\,h^{-1}\,M_\odot$. Using
Eq.~(\ref{eq:critlsize}), we measure the size of the critical curves
produced by our analytic model for different source redshifts. This is
repeated for all cosmological models we consider. For each of them, we
finally estimate the growth rate of the lens critical curves.

We show in Fig.~\ref{fig:grow} how the 
critical curve grows with source redshift. The curves
are normalised to the size of the critical curve for sources at
$z_{\rm s}=1$. Thin and thick lines show the results for ellipticities
$e=0$ and $e=0.3$, respectively. The growth of the critical curve
with redshift is larger in the $\Lambda$CDM
model than other cosmologies with dark energy. In this
case, for the axially symmetric lens, the growth between $z_{\rm s}=1$
and $z_{\rm s}=2$ is roughly by a factor of $4.5$. Beyond $z_{\rm
s}=2$, the growth slows down; for sources at
$z_{\rm s}=5$, critical curves are larger by a factor $\sim 7$ than
for sources at $z_{\rm s}=1$. Note that this increase is
determined not just by the larger distances to higher redshifts, 
but also by the steepening of the NFW profile at large radii which are
probed by the 
higher source redshifts. The model which deviates the most from
$\Lambda$CDM is the SUGRA model, for which the critical curves for
$z_{\rm s}=2$ and $z_{\rm s}=5$ are larger by factors of $\sim 3.6$
and $\sim 5.2$, respectively, than for $z_{\rm s}=1$. The RP and the
DECDM models fall between them.

Adding ellipticity to the model reduces the growth rate of the
critical curves for all cosmologies. For $e=0.3$, the critical curves
are larger by a factor of $\sim 3.4$ for $z_{\rm s}=2$ compared to
$z_{\rm s}=1$ in the $\Lambda$CDM model and by a factor of $\sim 2.8$
in the SUGRA model. The dependence of the growth rate on the lens
ellipticity will be discussed in detail in the following section.
 
\begin{figure}
  \includegraphics[width=\hsize]{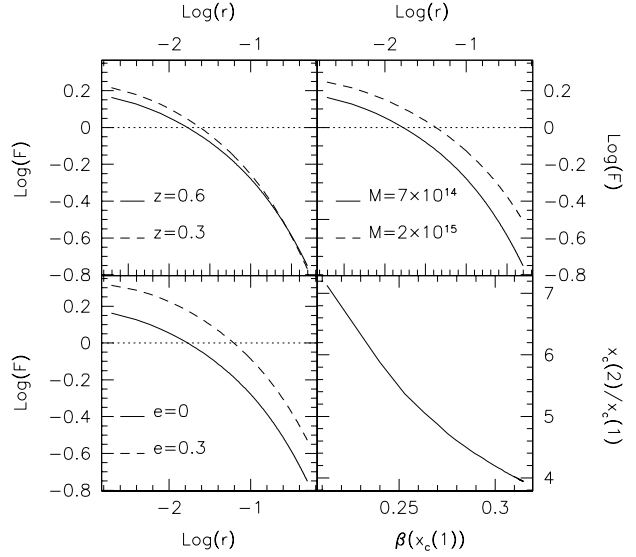}
\caption{Dependence of the
  critical-curve position on lens redshift, mass and
  ellipticity. Critical curves arise where $\log(F)=0$, as indicated
  by the horizontal dotted lines. Shown is $\log(F)$ as a function of
  $\log(r)$ for two lens redshifts (top left panel), for two lens
  masses (top right panel) and for two lens ellipticities (bottom left
  panel). In the bottom right panel we show the ratio of the
  critical-curve size for sources at redshift $z_{\rm s}=2$ and
  $z_{\rm s}=1$ as a function of the logarithmic slope $\beta$
  measured on the $z_{\rm s}=1$ critical curve.}
  \label{fig:kg_zme}
\end{figure}


\subsection{Dependence on lens mass, redshift and ellipticity}

\begin{figure*}
  \includegraphics[width=.33\hsize]{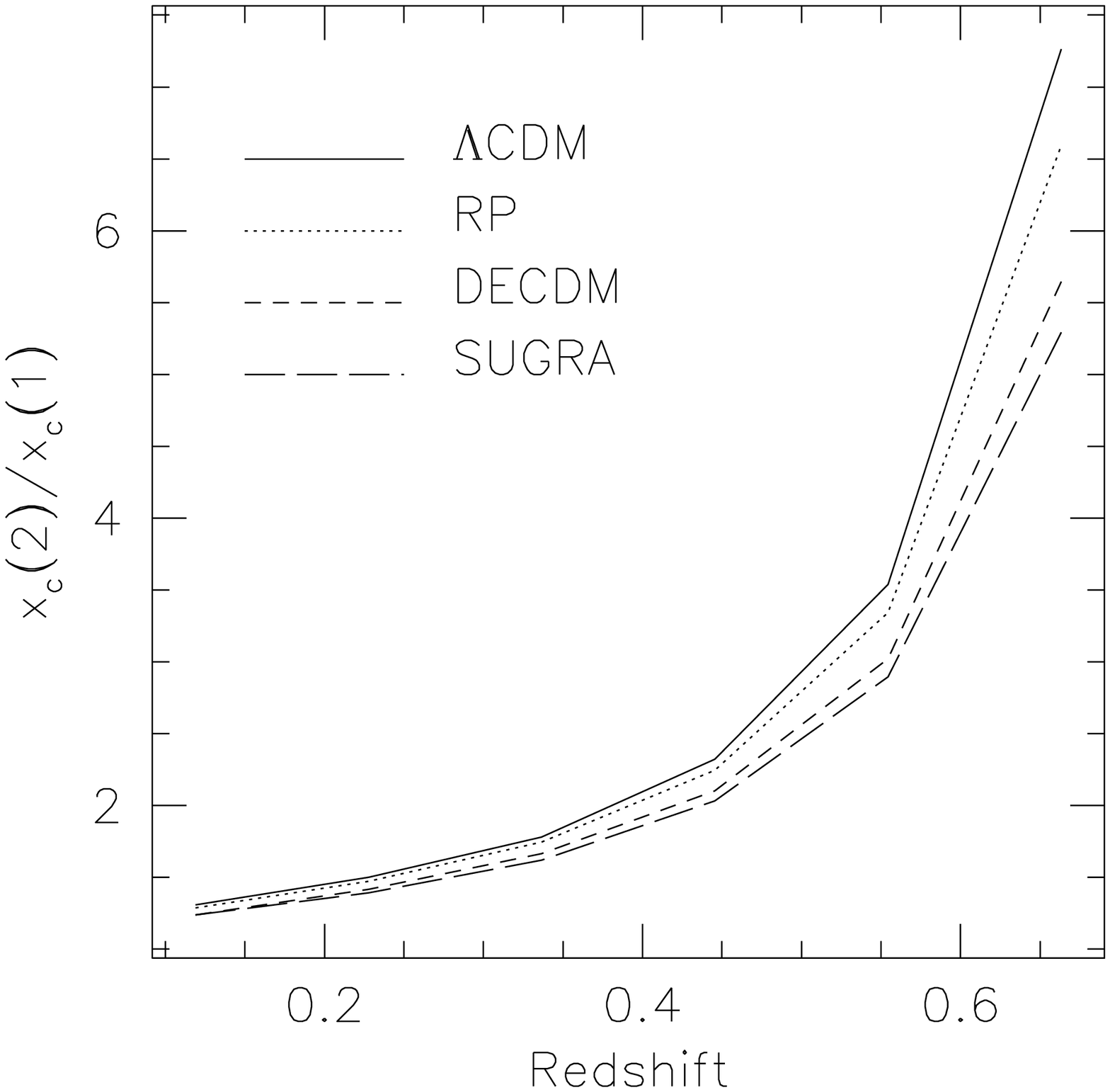}\hfill
  \includegraphics[width=.33\hsize]{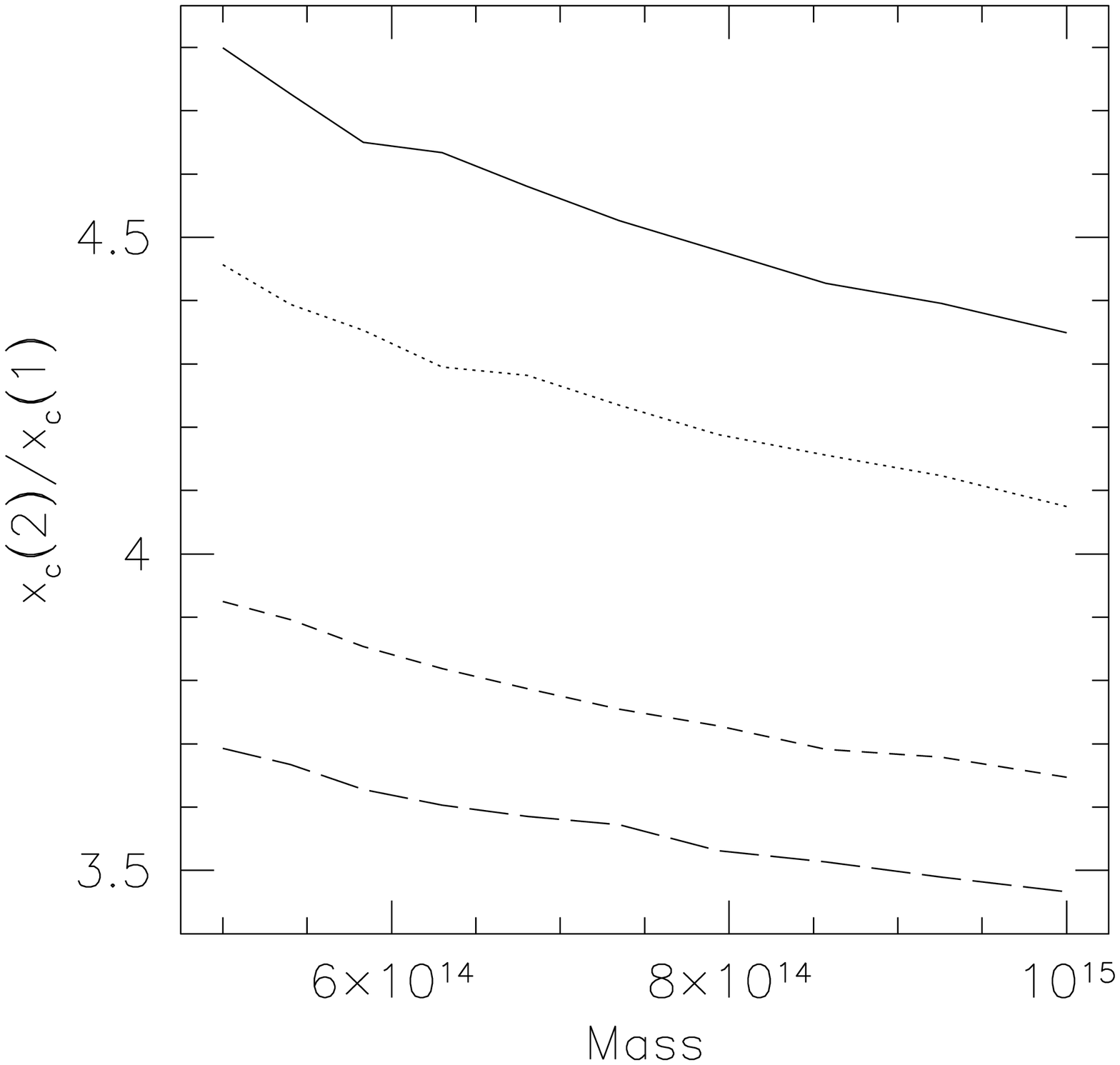}\hfill
  \includegraphics[width=.33\hsize]{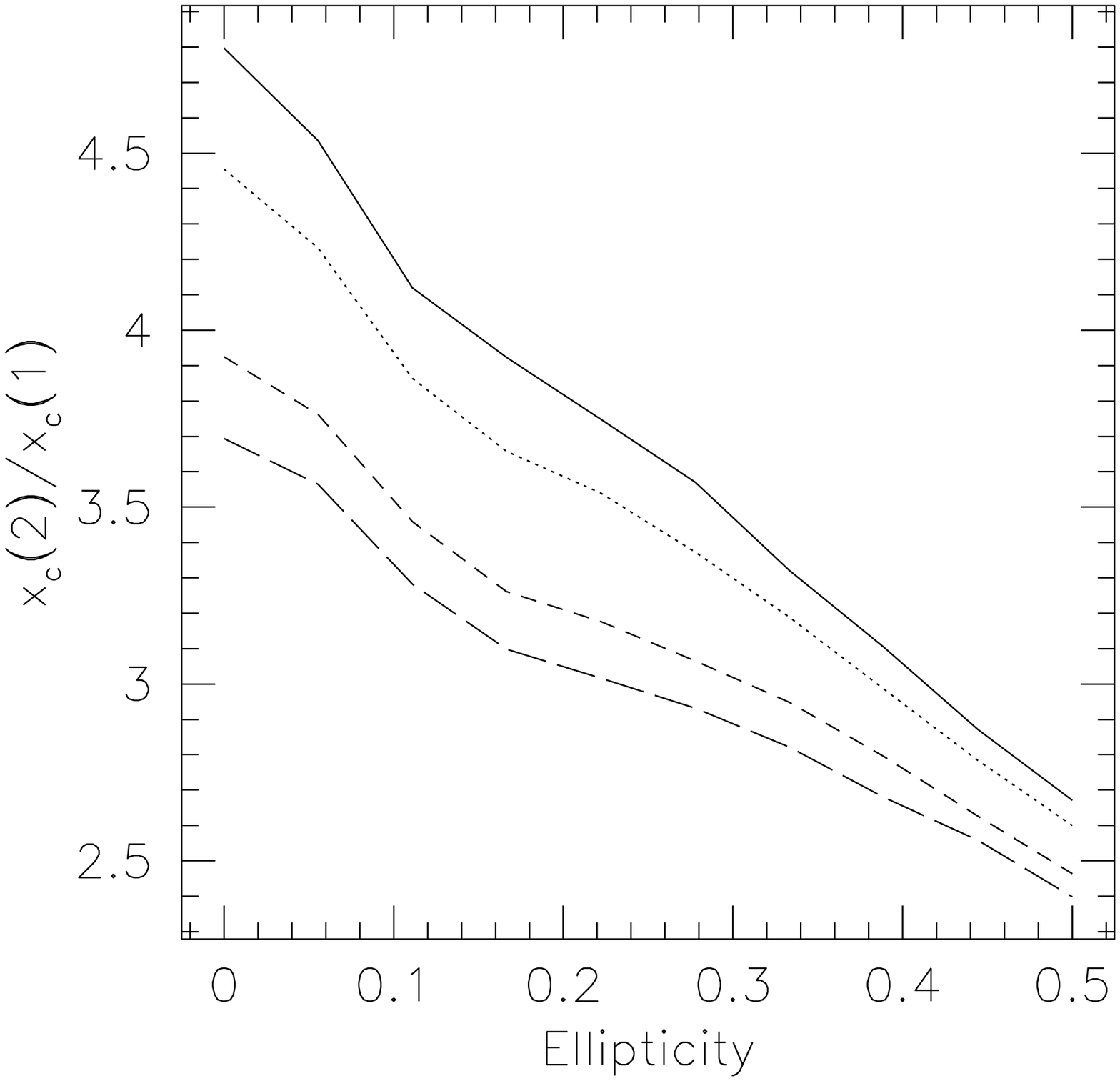}
\caption{Growth of the critical curve size as a function of the lens
  redshift, mass and ellipticity. 
  The ratio between the sizes of the critical curves at
  redshifts $z_{\rm s}=2$ and $z_{\rm s}=1$ is shown. }
  \label{fig:growzme}
\end{figure*}

As noted above, the amount by which the critical curves grow as a
function of source redshift depends on the lens ellipticity. In
this section, we discuss the dependence of the critical-curve 
growth rate on lens ellipticity, mass and redshift.

To explain the dependence on these lens properties, we must
consider where the critical curves form. In the case of an axially
symmetric model, they occur where the radial function
$F(r,z)=\kappa(r,z)+\gamma(r,z)$ is unity. For the NFW density
profile, $F(r,z)$ can locally be parametrised as
\begin{equation}
  F(r,z)\propto r^{-\beta(r,z)} \;,
\end{equation}
where the $\beta(r,z)$ is the logarithmic slope at radius $r$, 
\begin{equation}
  \beta(r,z)=-\left.
  \frac{\partial\log F(r,z)}{\partial\log r}\right|_r \;,
\end{equation}
which increases with radius.

The dependence of $F$ on the lens redshift, mass and ellipticity is
shown in the first three panels of Fig.~\ref{fig:kg_zme}. In each plot
we show $\log(F)$ as a function of $\log(r)$. Where the ellipticity is
varied, we plot $F$ along the major axis of the ellipse. When not
explicitely mentioned, the lens model has a mass of
$M=7\times10^{14}\,h^{-1}\,M_\odot$, a redshift of $z_{\rm l}=0.6$,
and an ellipticity of $e=0$.

Keeping the source redshift fixed at $z_{\rm s}=1$, changing the lens
redshift, mass or ellipticity causes the curve to be shifted up and
down, or left and right, implying that the critical curve forms at
radii characterised by different values of $\beta$. In particular,
when moving the lens closer to the sources or the observer, or
decreasing its mass and its ellipticity, $F$ reaches unity at larger
$\beta$.

The relative shift of the critical-curve position as a function of the
source redshift depends on the local slope of $\log(F)$ where the
critical curve forms. Since the tangential critical curve for sources
at redshift $z$ occurs where $F(r,z)=1$, the critical curve for
sources at redshift $z'=z+\d z$ lies where
\begin{equation}
  F(r',z')=F(r,z')+\frac{\partial F(r,z')}{\partial r}\d r=1 \, ,
\end{equation}
from which we obtain
\begin{equation}
  \frac{\partial F(r,z')}{\partial r}\Delta r=
  1-F(r,z')=\Delta F(r,z') \, .
\end{equation}
Since
\begin{equation}
  \frac{\partial F}{\partial r}=
  \frac{\partial \ln F}{\partial \ln r}
  \frac{F}{r}=\beta\frac{r}{F} \, , 
\end{equation}
the relative shift of the critical curve is given by
\begin{equation}
  \frac{\Delta r}{r}=\frac{1}{\beta} \frac{\Delta F}{F} \, .
\end{equation}
When $\log(F)$ is flatter, a larger relative shift of the critical
curve is obtained. This is shown in the last panel of
Fig.~\ref{fig:kg_zme}, where the ratio between the sizes of the
critical curves for sources at redshift $z_{\rm s}=2$ and $z_{\rm
s}=1$ is plotted as a function of $\beta$ measured at the position of
the critical curve for sources at $z_{\rm s}=1$. Here we have used as
a lens an axially symmetric model of mass
$M=7\times10^{14}\,h^{-1}\,M_\odot$ at redshift $z_{\rm l}=0.6$.

Since, as shown earlier, the value of $\beta$ on the critical curve
depends on the lens redshift, mass and ellipticity, we expect a
different growth rate of the critical curves for lenses with different
values of these parameters.

This is shown in Fig.~\ref{fig:growzme}. We plot
the ratio of the sizes of the critical curves for sources at
$z_{\rm s}=2$ and $z_{\rm s}=1$ as a function of lens redshift,
mass and ellipticity. These are shown for the four dark energy
models as indicated. As before, 
where not explicitly indicated, the lens model has
mass $M=7\times10^{14}\,h^{-1}\,M_\odot$, redshift $z_{\rm
l}=0.6$, and ellipticity $e=0$. As expected, the growth rate is
larger for lenses at higher redshift and with lower mass and
ellipticity. A stronger dependence is found on redshift and
ellipticity, while the dependence on mass is much weaker.

\begin{figure*}
  \includegraphics[width=.33\hsize,height=.35\vsize]{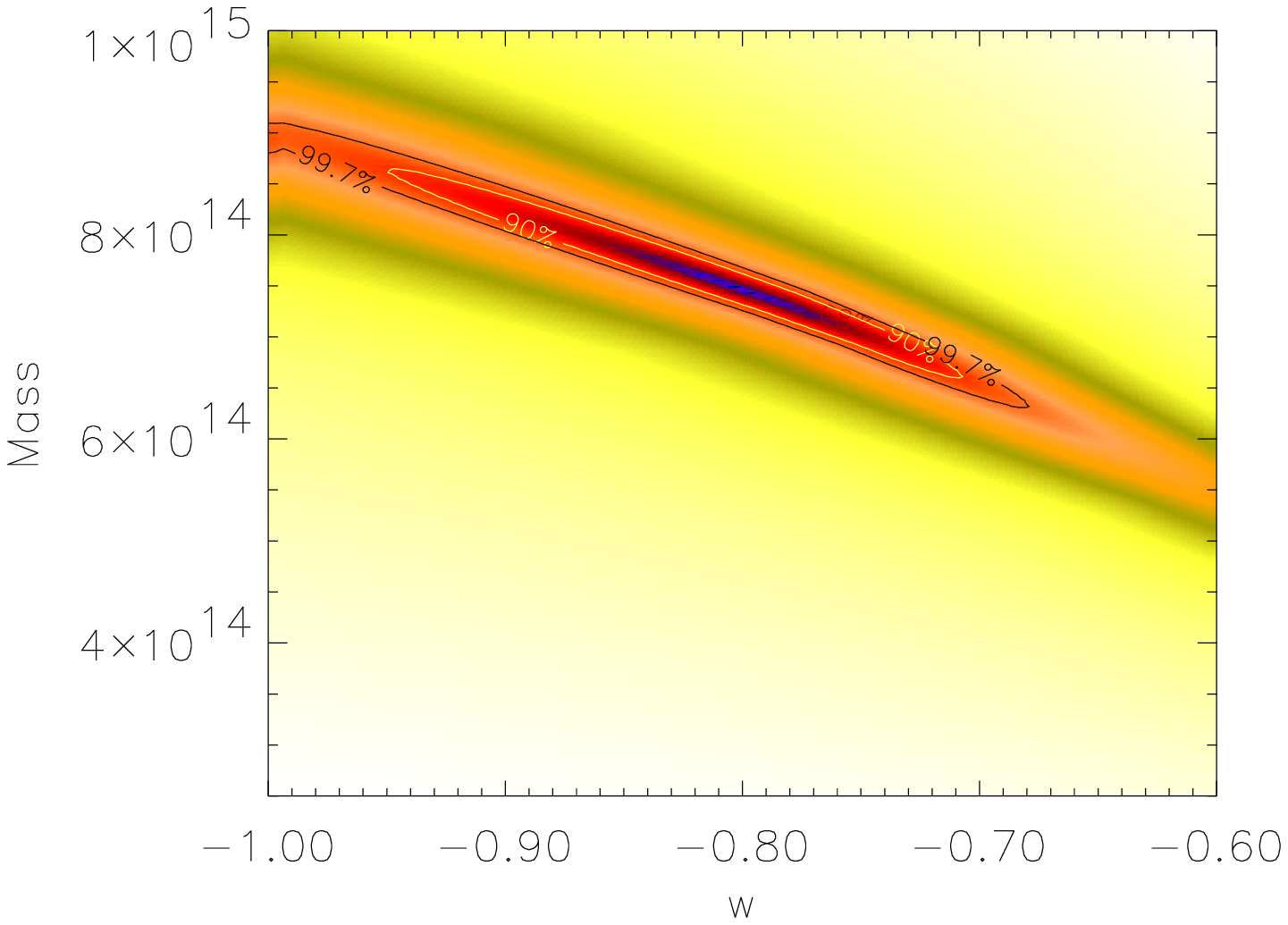}
  \includegraphics[width=.33\hsize,height=.35\vsize]{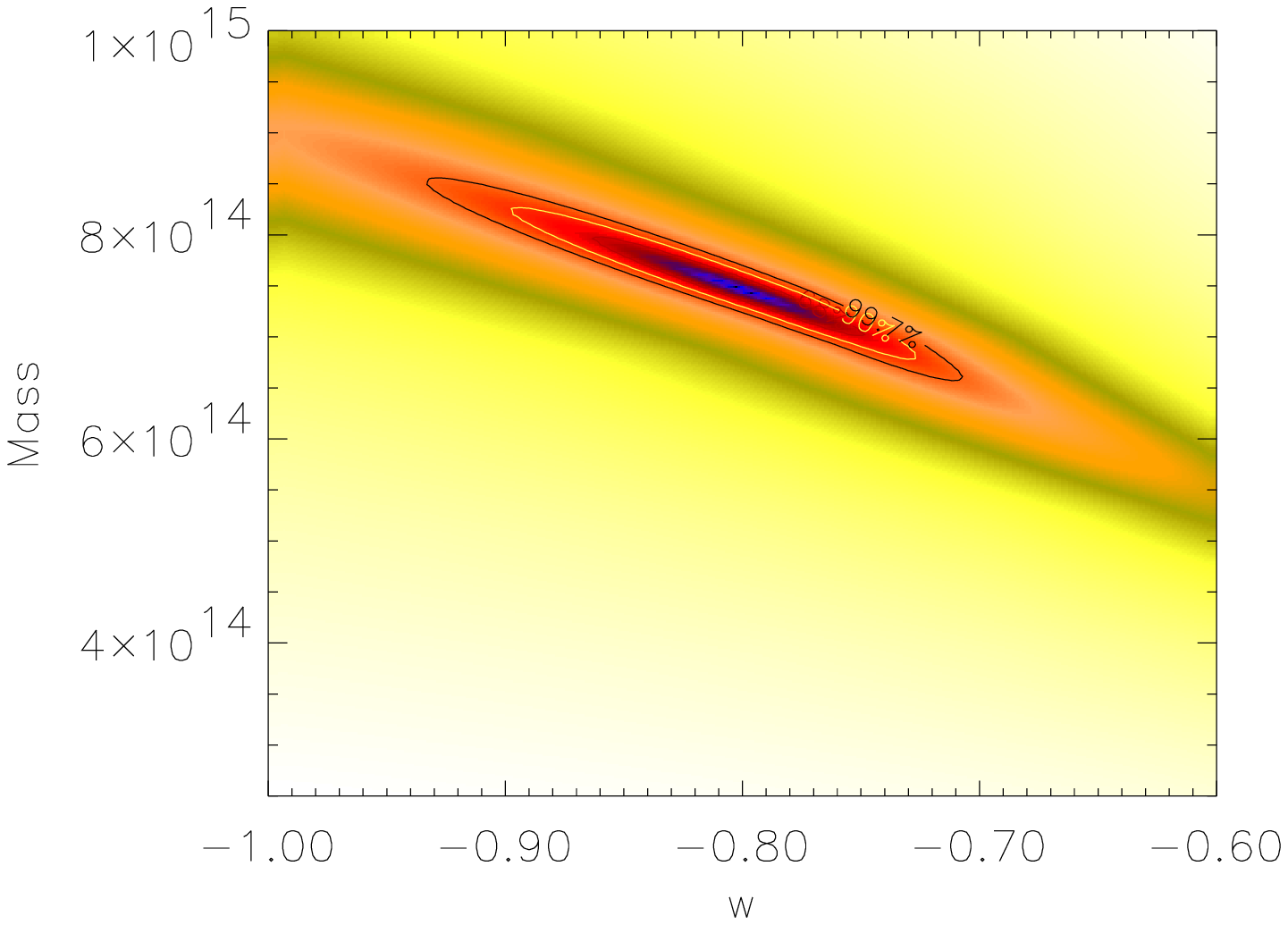}
  \includegraphics[width=.33\hsize,height=.35\vsize]{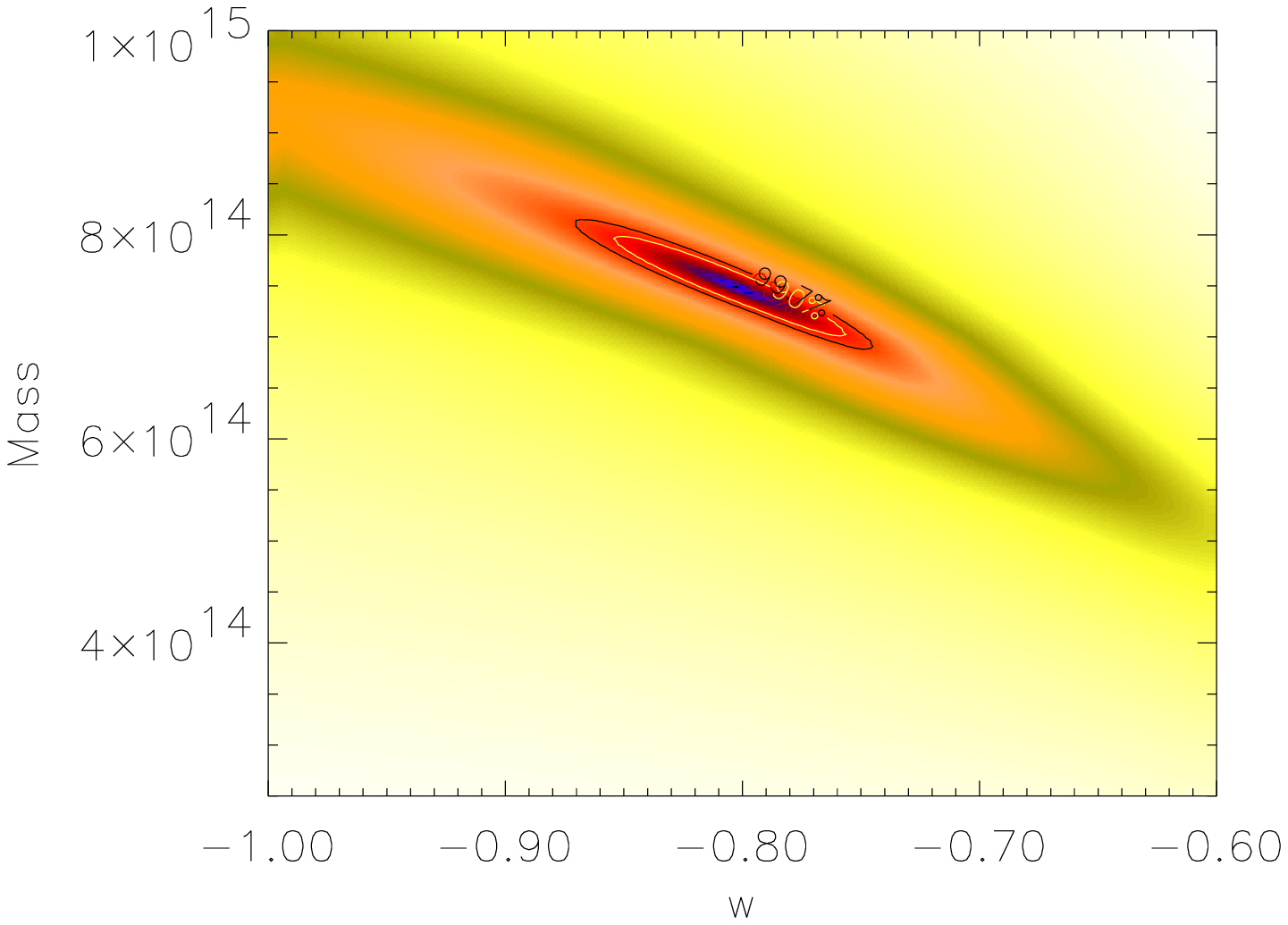}
\caption{Confidence levels in the $w$-mass plane. The input model is a
  lens with an NFW profile with mass $M=7.5\times
  10^{14}~h^{-1}M_{\odot}$ at $z=0.6$ in a cosmological model with
  $w=-0.8$.  The contours shown by the dark curve
  correspond to a probability level of $99.7\%$. 
  The three panels correspond to constraints from two arcs at
  $z=1$ and $z=2$ (left); from two arcs at $z=1$ and $z=2$ plus velocity
  dispersion at $10$ kpc from the centre (middle); from three arcs at
  $z=0.8$, $z=2$ and $z=3$ plus velocity dispersion at $10 kpc$ from the
  centre (right). Note that these constraints apply only for the
smooth analytical model we have used. }
  \label{fig:masswdeg}
\end{figure*}

We notice that the sensitivity of the displacement of the critical
curves to these lens properties changes as we vary the equation of
state of dark energy. This reflects the different evolution of halos
of a given mass in different cosmologies. As shown by
\cite{BA02.1,BA03.1,DO03.2}, the formation epoch of dark-matter halos
with the same mass in these cosmological models is significantly
different, leading to substantially different concentration parameters
of the halo density profiles. Halos tend to form earlier and thus have
typically larger concentrations in the SUGRA and in the DECDM models
compared to the RP and the $\Lambda$CDM models. Thus, for these
models, the growth rate of the critical curves is less sensitive to
changes in redshift, mass and ellipticity.

\subsection{Degeneracies in NFW models}

Since the expansion of the critical curves as a function of the source
redshift is different for halos with different concentrations, a
degeneracy between mass and cosmology arises in NFW halos.  That is, the same
displacement of the critical curves for sources at two redshifts can
occur with halos of different concentrations in more than one
cosmological models. As shown in Sect.~\ref{sect:genprop}, the
concentration is related to halo mass in NFW models of clusters.

For probing this degeneracy, we carry out the following test. We use
an input model consisting of a lens of mass $M=7.5 \times
10^{14}\,h^{-1}\,M_\odot$ at redshift $z_{\rm l}=0.6$ in a
cosmological model with $w=-0.8$. We consider its critical curves for
source redshifts $z_{\rm s}=1$ and $z_{\rm s}=2$, mimicking the
constraints from two tangential arcs, and we fit their positions by
varying the equation of state of dark energy and the lens mass. For
simplicity, we consider only cosmologies with time-independent
$w$. The fit is performed by minimising
\begin{equation}
  \chi_1^2=\left(
  \frac{x_{1}(M,w)-\hat{x}_{1}}{\Delta_{1}}\right)^2+\left(
  \frac{x_{2}(M,w)-\hat{x}_{2}}{\Delta_{2}}\right)^2\;, 
\end{equation}  
where $\hat{x}_1$ and $\hat{x}_2$ are the positions of the tangential
critical curves of the input model for sources at $z_{\rm s}=1$ and
$z_{\rm s}=2$, respectively, $\Delta_1$ and $\Delta_2$ are their
respective errors, and $x_1(M,w)$ and $x_2(M,w)$ are the corresponding
positions of the critical curves predicted by the fitting model with
mass $M$ in a cosmological model with dark-energy equation of state
$w$. We assume here to be in an idealized situation where the location of
the critical curves is known at the $1\%$ level.

We show the confidence levels in the $w$-$M$ plane resulting from this
fitting procedure in the upper left panel of
Fig.~\ref{fig:masswdeg}. The innermost and the outermost contours
correspond to probability levels of $68\%$ and $99.7\%$,
respectively. As anticipated, a good fit to the position of the
critical curves is obtained for a range of $M$ and $w$, with $99.7$
confidence limits ranging between $6\times10^{14}\,h^{-1}\,M_\odot
\lsim M \lsim 9\times10^{14}\,h^{-1}\,M_\odot$ and $-1\lsim w \lsim
-0.65$ (see also \citealt{CH02.2}).

For breaking this degeneracy, we must add constraints on the lens
density profile. One possibility is to use the stellar velocity
dispersion data for the brightest cluster galaxies in conjunction with
the arc positions and redshifts \citep[see][and references
therein]{MI95.1,SA03.1}. Including this additional constraint in our
fitting procedure, we re-define our $\chi^2$ variable as
\begin{equation}
  \chi^2_2=\chi^2_1+\left(\frac{\sigma_{r}(M,w)-\hat\sigma_{r}}
  {\Delta_{\sigma_r}}\right)^2\;, 
\end{equation}
where $\hat\sigma_{r}$ is the velocity dispersion of the input model
at $10\,h^{-1}$kpc from the centre, $\Delta_{\sigma_r}$ is the
uncertainty in its measurement, and $\sigma_{r}(M,w)$ the value
predicted by the fitting model. For calculating the velocity
dispersion from the density profile we use the spherical Jeans
equation, assuming isotropic orbits,
\begin{equation}
  \frac{\d \rho(r) \sigma_r(r)}{\d r}= -\frac{GM(r)\rho(r)}{r^2}\;,
\end{equation}
where $G$ is the gravitational constant and $M(r)$ is the mass
enclosed within a sphere of radius $r$. Again, we assume that the
velocity dispersion at the chosen radius is known with an accuracy of
$1\%$.  This is a rather optimistic assumption, but our primary aim
here is to discuss what kind of constraints are required for breaking
the degeneracy. The new map of the confidence levels is shown in the
upper right panel of Fig.~\ref{fig:masswdeg}. Although the contours
shrink, some degeneracy remains, indicating that further constraints
on the lens density profile are needed.

Finally, we use a lensed image from a third source at redshift $z_{\rm
s}=3$ for locating another critical curve, in addition to two at
$z_{\rm s}=0.8$ and $z_{\rm s}=2$ and to the velocity dispersion
constraints. This allows us to distinguish among different
cosmological models becomes easier as shown in the right
panel of Fig.~\ref{fig:masswdeg}, but does require that we get
lucky with the arc redshifts. It is also valid only for the smooth
mass distribution represented by our analytical model; real clusters
are likely to have a more lumpy structure. 

Our simplified analysis with analytical NFW models for cluster halos
was aimed at understanding the sensitivity of arc locations to
variations in several physical parameters. Our results show that to
get robust constraints on cosmology from a single cluster, a detailed
modelling of the lensing mass distribution is required (see also
\citealt{CH02.2,DA03.1}). This modelling would include the
contribution from sub-structure in the cluster which we have not
studied in our analytical models. Further, as shown by \cite{DA04.1},
structures along the line of sight provide a source of error in
modelling individual clusters that is very difficult to overcome.

\section{Numerical models}

\subsection{Cluster sample}

Since asymmetries and substructures play a crucial role in determining
the strong lensing properties of galaxy clusters \citep{ME03.1},
analytic models can only be used for an approximate description of
their lensing properties. More realistic mass distributions of
clusters, as provided by numerical simulations, are needed for drawing
quantitative conclusions. We now repeat the analysis previously
applied to analytic models to a sample of numerical clusters.

The sample of clusters we use here consists of the 17 halos used by
\cite{DO03.2} and \cite{ME04.1}. We briefly summarise their main
properties. The cluster models are pure dark-matter halos and were
obtained using the most recent version of the cosmological code GADGET
\citep{SP01.1}. The code was extended by \citet{DO03.2} to
cosmological models with dynamical dark energy. Each cluster was
obtained by resimulating at higher resolution a patch of a
pre-existing large-scale cosmological simulation \citep{TO97.2}. The
initial conditions were set up with the purpose of obtaining
identically comparable clusters in all the cosmologies at redshift
$z=0$. Of course, at higher redshifts, each of them appears at
different evolutionary stages in different cosmologies, since the
growth of the density perturbations depends on the equation of state
of the dark energy. For example, clusters form earlier in the SUGRA
and the DECDM compared to the RP and the $\Lambda$CDM models. The
implications of the different time evolution of these clusters for
different equations of state of dark energy on the abundance of strong
lensing events is discussed in detail by \cite{ME04.1}.

The clusters in the sample contain on average $N_V\approx200,000$ dark
matter particles within their virial radii. The corresponding virial
masses range between $M_V=3.1\times10^{14}$ to
$1.7\times10^{15}\,h^{-1}\,M_\odot$ at redshift zero.

\subsection{Lensing simulations}

\begin{figure*}
  \includegraphics[width=\hsize]{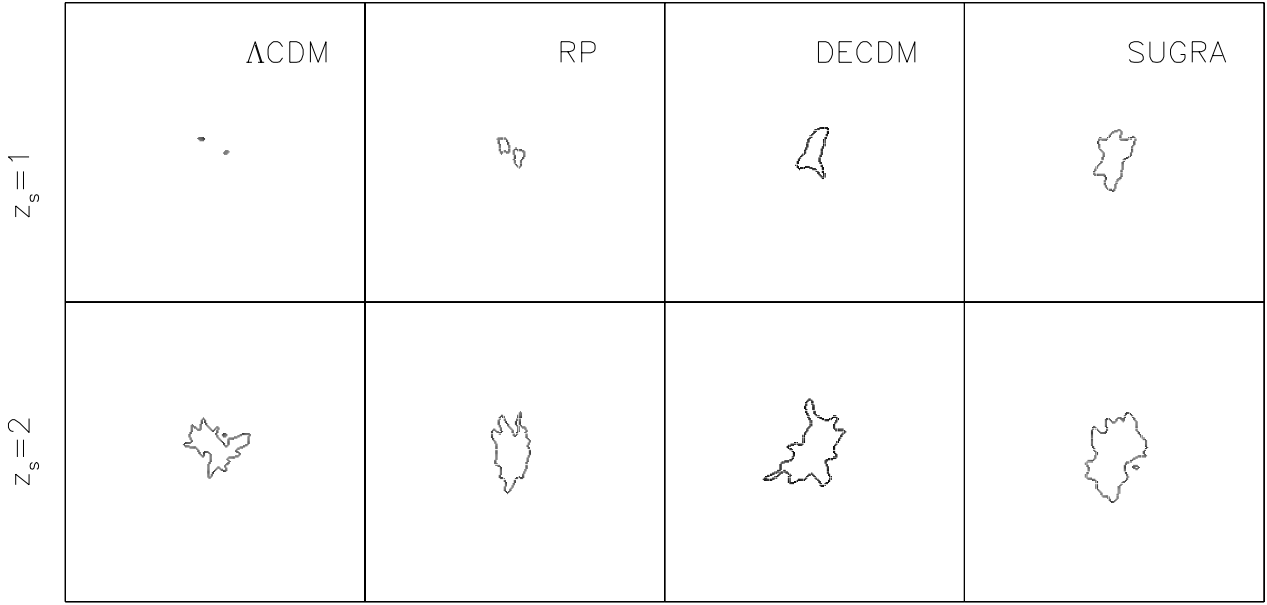}
\caption{Example showing how the critical curves of a numerical
  cluster at $z_{\rm l}=0.6$ simulated in different cosmological
  models change between $z_{\rm s}=1$ and $z_{\rm s}=2$}
  \label{fig:numcc}
\end{figure*}

For each of them we take the snapshot at redshift $z_{\rm l}=0.6$ and
perform lensing simulations. We first select from the simulation
box a cube of comoving side length $3 \, h^{-1}$Mpc. The
three-dimensional mass distribution of the cluster is projected on a
regular grid of $256 \times 256$ cells along three orthogonal
axes. The resulting two-dimensional density fields are smoothed using
the {\em Triangular Shaped Cloud} method \citep{HO88.1} for avoiding
discontinuities between neighbouring cells. Thus, we analyse 51
projected lens mass distributions.

Through each of the surface density maps, we trace a bundle of $1024
\times 1024$ light rays on a regular grid covering the central $1 \,
h^{-1}$Mpc$^2$ of the lens plane. This is large enough for enclosing
the critical curves of all our numerical models. Deflection angles are
computed using the method described in \citet{ME00.1}. We first define
a grid of $128\times128$ ``test'' rays, for each of which the
deflection angle is calculated by directly summing the contributions
from all cells on the surface density map $\Sigma_{i,j}$,
\begin{equation}
  \vec{\hat{\alpha}}_{h,k}=\frac{4G}{c^2}\sum_{i,j} \Sigma_{i,j} A
  \frac{\vec x_{h,k}-\vec x_{i,j}}{|\vec x_{h,k}-\vec x_{i,j}|^2}\;,
\end{equation}  
where $A$ is the area of one pixel on the surface density map and
$\vec x_{h,k}$ and $\vec x_{i,j}$ are the positions on the lens plane
of the ``test'' ray ($h,k$) and of the surface density element
($i,j$). Following \cite{WA98.2}, we avoid the divergence when the
distance between a light ray and the density grid-point is zero by
shifting the ``test'' ray grid by half-cells in both directions with
respect to the grid on which the surface density is given. We then
determine the deflection angle of each of the $1024\times1024$ light
rays by bi-cubic interpolation between the four nearest test rays.

The {\em reduced} deflection angle is
\begin{equation}
  \vec\alpha(\vec x)=
  \frac{D_{\rm ls}}{D_{\rm s}} \vec{\hat{\alpha}}(\vec x)\, .
\end{equation}
Since we wish to estimate the growth of the tangential critical curves
as a function of source redshift, we consider sources between $z_{\rm
s}=1$ and $z_{\rm s}=2$.

Since the reduced deflection angle is the gradient of the lensing
potential,
\begin{equation}
  \vec\nabla \psi(\vec x)=\vec{\alpha}(\vec x) \, ,
\label{eq:potang}
\end{equation}
the convergence and the shear can be easily written as:
\begin{eqnarray}
  \kappa(\vec{x})&=&\frac{1}{2}\left(\frac{\partial \alpha_{1}}{\partial
  x_1}+\frac{\partial \alpha_{2}}{\partial x_2}\right) \\
  \gamma_1(\vec{x}) &=& \frac{1}{2}\left(\frac{\partial \alpha_{1}}{\partial
  x_1}-\frac{\partial \alpha_{2}}{\partial x_2}\right) \\
  \gamma_2(\vec{x}) &=& -\frac{\partial \alpha_1}{\partial x_2}=
  -\frac{\partial \alpha_2}{\partial x_1} \ . 
\end{eqnarray}
The tangential critical curves are then determined by searching for
those points in the lens plane where Eq.~(\ref{eq:lambdat}) is
satisfied.

\subsection{Critical-curve sizes}

The critical curves for one of the clusters in our sample in the
different cosmological models are shown in Fig.~\ref{fig:numcc}. 
The sources are at redshift $z_{\rm s}=1$ in the upper panels and
at $z_{\rm s}=2$ in the lower panels. Two
important features are evident. First, for sources at redshift $z_{\rm
s}$ the sizes of the critical curves differ substantially among the
various cosmological models. For example, the cluster has almost no
critical curves in the $\Lambda$CDM model, while they are already well
developed in the SUGRA model. The RP and the DECDM models fall between
these two cosmologies. This is a consequence of the earlier formation
epoch of clusters in the RP, DECDM and in the SUGRA models than in the
$\Lambda$CDM model \citep[see e.g.][]{BA02.1,BA03.1,DO04.1,ME04.1},
due to which they have a larger concentration enabling them to be
efficient lenses even at relatively high redshifts or for relatively
close sources. Second, as expected from the analytical calculations,
the relative enlargement of the critical curves is higher in the
$\Lambda$CDM than in the other cosmological models.

\begin{figure}
  \includegraphics[width=\hsize]{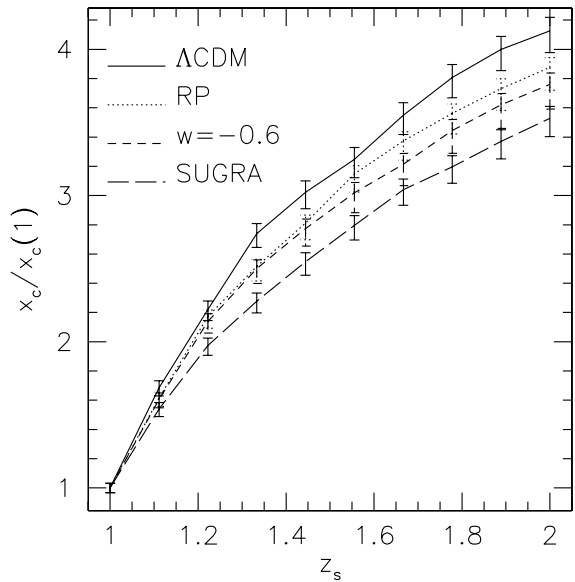}
\caption{Critical curve sizes normalized to $z_s=1$, as in 
  Fig.~\ref{fig:grow}, but for a numerical cluster
  sample comprising 17 halos placed at redshift $z_{\rm l}=0.6$. 
  Only clusters producing critical curves for $z_{\rm s}=1$ are
  considered. Curves display the sample medians. The error bars 
  show the first and the third quartile of the
  distribution, rescaled assuming that the information
  from $1000$ pairs of arcs could be combined.}
  \label{fig:medians}
\end{figure}

We apply the same method used for the analytical models for estimating
the size of critical curves of the numerical clusters. 
Fig.~\ref{fig:medians} shows the relative growth of the
critical curves in the four cosmological models as measured in the
numerical simulations. Each curve represents the median among the 51
halos which develop a critical curve for source redshift $z_{\rm
s}=1$. The number of useful clusters for this analysis ranges between
$\sim 20$ in the $\Lambda$CDM to $\sim 30$ in the SUGRA model. The
results confirm the qualitative expectations from analytic models: 
namely, the trend for different cosmologies. 
The absolute values of the relative growth are also consistent with the
predictions for a moderate ellipticity ($e \lsim 0.3$) lensing
potential (compare Figures 2 and 7).

These results show that the statistical application of this method is
potentially powerful. Upcoming surveys from space, like those which
will be conducted by SNAP \citep{AL04.1}, could provide detailed
observations of order thousand galaxy clusters, allowing the
information from many lenses to be combined. The error bars in
Fig.~\ref{fig:medians} show the first and the third quartiles of the
curve distribution we obtain from our numerical cluster sample. They
were rescaled to the expected error when the information from $\sim
1000$ pairs of arcs is combined. The figure shows that when
constraints on the position of the critical curves from sources at
significantly different redshifts and in a sufficiently large sample
of clusters are used, it becomes possible to discriminate among
different cosmological models. We have used a simple choice of source
redshifts; by extending the analysis to a wider redshift range,
especially to redshifts beyond 2 for the distant arcs, the constraints
became stronger. The analysis needs to be extended in other ways as well, 
by combining the information from different lens redshifts and 
finding the best way to weight a given cluster. 

In our test, we have used a simple estimate of the size of the
critical curves, which we can assume is related to the
cluster-centric distance of an observed arc, without any mass
modelling of the cluster. We obtain
substantially different amounts of growth of the critical curves
depending on the level of substructure in the central region of the
clusters. If the cluster is close to critical for sources at redshift
$\sim 1$ at the location of a substructure, we measure a rapid growth
of the critical curve once the sources are shifted to higher redshift. 
This occurs because the critical surface density becomes smaller and 
the cluster critical curves wiggle around the emerging critial mass 
lumps. This is shown for example in Fig.~\ref{fig:numcc}, where the 
shape of the critical curves changes dramatically between $z_{\rm
s}=1$ and $z_{\rm
s}=2$ for some of the cosmological models (see e.g.~the
DECDM model). If mass modelling for some of the clusters were to be
included, the scatter in Fig.~\ref{fig:medians} could be reduced.
Finally we note that to some extent carrying
out such an analysis from observations would require simulations that
correctly reproduce the properties of real clusters. While we have
used only the relative locations of arcs at different redshifts, it
is worth examining what aspects of cluster structure these are
sensitive to.

\subsection{Imaging surveys for cluster lensing}

Strong lensing studies with galaxy clusters require high quality
imaging data. While we will not examine the survey requirements 
in any detail, we have used our numerical clusters to produce 
lensed images of the Hubble Ultra Deep Field (HUDF). 
We can use these to study of the effects of sky brightness, 
photon noise and resolution on cluster arcs. 

Figure 8 shows the kind of imaging expected
from space imaging and current state of the art ground based
imaging. Clearly, the demands on resolution are very high to reliably
measure the properties of lensed arcs. With $0.5$ arcsecond imaging 
the quality of the
images is compromised, as shown in the right panel, but the large
arcs are still identified. Adaptive optics may enable better 
resolution images to be obtained than shown in this figure; short 
of that, it may still be possible to use giant arcs and on average
get the arc separations accurately enough. Ideally, a space based 
survey that images a large number of multiple-arc 
clusters with resolution comparable to the HST would provide an
adequate sample. For such a survey, the question of 
how deep it is worth going to find
high-redshift arcs merits worth further study. 

In addition to multi-color imaging, which would enable photometric
redshift estimates, cluster cosmography would require spectroscopic
redshifts for the lensed arcs.  An important exercise for future work
is to estimate the number of arcs expected from a realistic sample of
clusters for given survey parameters. \cite{DA03.1} have used three
existing cluster lensing datasets \citep{GL03.1,LU99.1,ZA03.1} and
estimated that of order 1000 arcs are expected on the sky even from a
survey of modest depth.  Given the landscape of proposed telescopes
and surveys, it is not prohibitive to think high resolution imaging of
order a thousand cluster containing fields, as well as follow-up
spectroscopy on the several thousand arcs that might be found.

\begin{figure*}
  \includegraphics[width=\hsize]{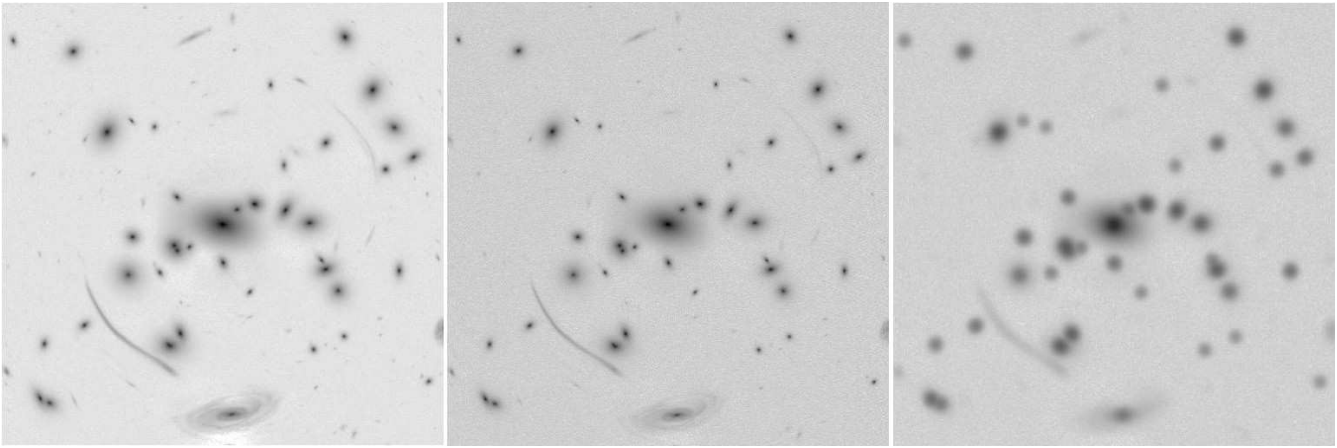}
\caption{Simulated observations from a VLT-like telescope of the same field
observed by HST in the ultra-deep field with $10^6$ sec exposure. The
simulations include a sky brightness $I=19.9$ mag/arcsec$^2$, Poisson
photon noise and a seeing of $0.5$ arcsec. The lensing cluster of
galaxies is at $z=0.25$. The cluster members have been added by placing
a cD galaxy at the cluster center and by distributing the other
galaxies following the underlying dark matter mass distribution. The
galaxy luminosity reflects that of realistic cluster members (spectral
energy distribution, morphological type as a function of the the
cluster-centric distance reproduce observational constraints). The
integration time is 3000 sec. In the lensing simulation, all the
sources were assumed to be at $z=2$. The field size is $~1$ arcminute 
along the x-axis. The middle panel includes sky brightness and Poisson
photon noise; the right panel includes seeing. 
}
\label{fig:hudf}
\end{figure*}
 
\section{Summary}

This paper presents a study of cosmography with galaxy clusters that
produce lensed arcs of background galaxies. Clusters with multiple
arcs from different redshifts provide constraints on angular diameter
distances at different redshifts. These in turn constrain the equation
of state of the dark energy. We have studied this effect with
analytical models to understand some of the qualitative trends, 
and with numerical simulations to see what can be expected from
surveys of real clusters. 

We have used analytical models of clusters based on the NFW profile to
study how the positions of arcs depend on the mass distribution and
the cosmological model. We find that the ellipticity and concentration
of the cluster halo can shift the critical curves. Substructure would
have the same effect. This means that for individual clusters, it is
important to have independent information on the mass distribution,
especially inside the critical curve, to extract cosmographic
information from arc positions and redshifts. Our analysis overlaps
with other recent studies of cluster arcs,
e.g.~\cite{CH02.2,OG03.1,ME03.1,DA03.1,WA03.1}. We explore examples of
how information from inner velocity dispersions and multiple arcs
could be combined for a single cluster. However a more sophisticated
study is warranted.

Given the limitations of analytical models (they cannot for example
include the effect of substructure in the halos), we conclude that
numerical modelling is essential to extract information from observed
clusters. We use a sample of simulated clusters in four dark energy
models in an exploratory study of whether future surveys can provide
cosmographic information. We examine the scatter in the relative
positions of arcs due to differences in the mass distributions of
individual clusters.  We focus on the critical curves and use these to
estimate arc positions. While this is a simplified study, we believe
it gives us a reasonable estimate of the minimum requirements of a
cluster sample to overcome the cluster-to-cluster scatter and obtain
cosmographic constraints.

In the numerical study we have not attempted to model the mass
distribution of individual clusters. We have only used information on
the relative sizes of critical curves at different redshifts. In
practice, some constraints on the mass distributions will be obtained
from strong and weak lensing, as well as X-Ray, SZ and velocity
dispersion measurement. Further study is needed on how well this
information can be used to improve cosmographic constraints.  Further,
we have used information only from arcs up to redshift 2; one would do
better if imaging and spectroscopy on arcs at higher redshifts were
available.

We have attempted to include the main sources of complexity that arise
from the mass distribution of clusters. We have neglected several
factors in the observation of clusters and their interpretation that
merit further study. Structures along the line of sight at redshifts
different from the cluster are an important source of
scatter. \cite{DA04.1} have studied the impact of these structures;
they find that they limit the use of individual clusters and could
contribute to the error in dark energy parameters from an ensemble of
clusters.  Further we used just the positions of the largest arcs, and
have largely ignored other strong lensing information as well as the
measurement error in using arc positions. These factors are all
important and require more detailed studies.

We have estimated the scatter in cosmographic constraints from
different clusters. We have not addressed the question of how an
actual survey would best average results from different
clusters. Clearly some clusters  will have multiple
arcs (over 100 in the case of Abell 1689) and therefore contain more
information. Others may have a regular, spherically symmetric
structure which would make it easier to extract cosmographic
information from arc positions. 

\bigskip
While completing this study, we have learnt of a preprint by
\cite{DA04.1} that examines sources of noise in arc-cosmography. 

\section*{Acknowledgements}

We are grateful to Francesca Perrotta and Carlo Baccigalupi for their
substantial input and helpful comments on the paper. We thank Neal 
Dalal, Gary Bernstein, Mike Jarvis, Phil Marshall, 
Yannick Mellier, David Rusin, 
and Masahiro Takada for helpful discussions. B.J. is supported in part 
by NASA grant NAG5-10924 and the Keck foundation. 

\bibliography{}
\bibliographystyle{mn2e}

\end{document}